\documentclass[fleqn,usenatbib]{mnras}
\usepackage[T1]{fontenc}
\usepackage{graphicx}	
\usepackage{amsmath}
\usepackage{amssymb}
\usepackage{ae,aecompl}
\usepackage{newtxtext,newtxmath}
\usepackage{soul}
\usepackage{comment}

\DeclareRobustCommand{\VAN}[3]{#2}
\let\VANthebibliography\thebibliography
\def\thebibliography{\DeclareRobustCommand{\VAN}[3]{##3}\VANthebibliography}

\title[Dark-matter-deficient dwarf galaxies]{
Dark-matter-deficient dwarf galaxies form via tidal stripping of dark matter in interactions with massive companions}

\author[R. A. Jackson et al.]
{R. A. Jackson,$^{1}$\thanks{E-mail: r.jackson9@herts.ac.uk}
S. Kaviraj,$^{1}$ G. Martin,$^{2,3}$, J. E. G. Devriendt,$^{4}$ A. Slyz,$^{4}$ J. Silk,$^{4,5,6}$ \newauthor 
Y. Dubois,$^{5}$
S. K. Yi,$^{7}$
C. Pichon,$^{5,8}$
M. Volonteri,$^{5}$
H. Choi,$^{7}$
T. Kimm,$^{7}$
K. Kraljic,$^{9}$ and \newauthor
S. Peirani$^{5,10}$ 
\\
$^{1}$Centre for Astrophysics Research, School of Physics, Astronomy and Mathematics, University of Hertfordshire, Hatfield, AL10 9AB, UK\\
$^{2}$Steward Observatory, University of Arizona, 933 N. Cherry Ave, Tucson, AZ 85719, USA\\
$^{3}$Korea Astronomy and Space Science Institute, 776 Daedeokdae-ro, Yuseong-gu, Daejeon 34055, Korea\\
$^{4}$Dept of Physics, University of Oxford, Keble Road, Oxford OX1 3RH UK\\
$^{5}$Institut d'Astrophysique de Paris, Sorbonne Universit\'es, UMPC Univ Paris 06 et CNRS, UMP 7095, 98 bis bd Arago, 75014 Paris, France\\
$^{6}$Department of Physics \& Astronomy, The Johns Hopkins University, Baltimore, MD 21218, USA\\
$^{7}$Department of Astronomy and Yonsei University Observatory, Yonsei University, Seoul 03722, Republic of Korea\\
$^{8}$School of Physics, Korea Institute for Advanced Study (KIAS), 85 Hoegiro, Dongdaemun-gu, Seoul, 02455, Republic of Korea\\
$^{9}$Institute for Astronomy, Royal Observatory, Edinburgh EH9 3HJ, UK\\
$^{10}$Observatoire de la C$\hat{\rm{o}}$te d'Azur, CNRS, Laboratoire Lagrange, Bd de l'Observatoire, Universit\'e C$\hat{\rm{o}}$te d'Azur, CS 34229, 06304 Nice Cedex 4, France
}

\begin{document}
\label{firstpage}
\pagerange{\pageref{firstpage}--\pageref{lastpage}}
\maketitle

\begin{abstract}
In the standard $\Lambda$CDM paradigm, dwarf galaxies are expected to be dark-matter-rich, as baryonic feedback is thought to quickly drive gas out of their shallow potential wells and quench star formation at early epochs. Recent observations of local dwarfs with extremely low dark matter content appear to contradict this picture, potentially bringing the validity of the standard model into question. We use \texttt{\texttt{NewHorizon}}, a high-resolution cosmological simulation, to demonstrate that sustained stripping of dark matter, in tidal interactions between a massive galaxy and a dwarf satellite, naturally produces dwarfs that are dark-matter-deficient, even though their initial dark matter fractions are normal. The process of dark matter stripping is responsible for the large scatter in the halo-to-stellar mass relation in the dwarf regime. The degree of stripping is driven by the closeness of the orbit of the dwarf around its massive companion and, in extreme cases, produces dwarfs with halo-to-stellar mass ratios as low as unity, consistent with the findings of recent observational studies. $\sim$30 per cent of dwarfs show some deviation from normal dark matter fractions due to dark matter stripping, with 10 per cent showing high levels of dark matter deficiency (M$_{\rm{halo}}$/M$_{\star}$<10). Given their close orbits, a significant fraction of dark-matter-deficient dwarfs merge with their massive companions (e.g. $\sim$70 per cent merge over timescales of $\sim$3.5 Gyrs), with the dark-matter-deficient population being constantly replenished by new interactions between dwarfs and massive companions. The creation of these galaxies is, therefore, a natural by-product of galaxy evolution and their existence is not in tension with the standard paradigm.
\end{abstract}



\section{Introduction}
\label{sec:Intro}
In the standard $\Lambda$CDM paradigm, dwarf galaxies (M$_{\star}$ < 10$^{9.5}$ M$_{\odot}$) are expected to be dark matter (DM) rich, because their shallow potential wells make it easier for processes like stellar and supernova feedback to drive gas out from their central regions at early epochs. This reduces their star formation rates and produces systems that have a lower stellar content at a given DM halo mass \citep{Dubois2008,DiCintio2017,Chan2018,Jackson2020}. Dwarf galaxies are expected to form a natural extension to the stellar-to-halo mass relation in massive galaxies \citep{Moster2010,Read2017}, with the  M$_{\rm{halo}}$/M$_{\star}$ values expected to progressively increase towards lower stellar masses.  
However, some recent observational studies appear to challenge this picture. A growing number of studies suggest that some dwarf galaxies may deviate strongly from the expected stellar-to-halo mass relation \citep[e.g.][]{vanDokkum2018,vanDokkum2019,Guo2020,Hammer2020}, with unexpectedly low DM fractions. For example, \citet{Guo2020} have found 19 nearby dwarf galaxies that are DM deficient. The majority of these galaxies (14/19) appear to be relatively isolated, without many nearby bright galaxies, implying that they might have been born DM deficient. \citet{vanDokkum2018,vanDokkum2019} have studied two dwarf galaxies in group environments which could exhibit M$_{\rm{halo}}$/M$_{\star}$ values (within 7.6 kpc) close to unity, suggesting that they may have DM fractions that are around 400 times lower than that expected for galaxies of their stellar mass. There is still some controversy surrounding the validity of these findings \citep[e.g.][]{Oman2016,MartinN2018,Blakeslee2018,Emsellem2019,Fensch2019,Trujillo2019}, mostly focusing on the reliability of obtaining accurate distance measurements for these systems. For example, in the case of the \citet{vanDokkum2018} galaxy, a measured distance of 13 Mpc, as suggested by \citet{Trujillo2019}, rather than the quoted $\sim$ 20 Mpc in \citet{vanDokkum2018}, would make M$_{\rm{halo}}$/M$_{\star}$ $>$ 20, making it a relatively normal dwarf galaxy (but see \citet{vanDokkum2018b} for an alternate view). The existence of galaxies that are deficient in DM could pose a serious challenge to the $\Lambda$CDM model, as it is difficult for galaxies that are rich in baryons to form in halos that are deficient in DM. It is, therefore, important to investigate whether there are natural channels for the formation of such galaxies in the standard model. 

Some formation methods for DM deficient galaxies have been suggested in the recent literature. A well-known channel for forming such systems are tidal dwarfs \citep{Wetzstein2007,Bournaud2008b,Bournaud2008a,Kaviraj2012,Kroupa2012,Ploeckinger2015,Haslbauer2019}. These dwarf galaxies are formed in the tidal tails that emerge as a result of gas-rich major mergers of massive galaxies, either through Jeans instabilities within the gas which lead to gravitational collapse and the formation of self-bound objects \citep{Elmegreen1993}, or a large fraction of the stellar material in the progenitor disk being ejected and providing a local potential well into which gas condenses and fuels star formation \citep[e.g.][]{Barnes1992b,Duc2004,Hancock2009}. However, the contribution of tidal dwarfs to the galaxy population, particularly at the stellar masses of the DM deficient galaxies found by recent observational studies ($\sim$10$^9$ M$_{\odot}$), is extremely small \cite[e.g.][]{Kaviraj2012}. 

It has been postulated that DM deficient galaxies could form via high velocity, gas-rich mergers of dwarf galaxies themselves \citep{Silk2019,Shin2020}. In this scenario, DM deficient galaxies form as these mergers separate DM from the warm disc gas, which is then compressed by tidal interactions and shocks to form stars. Another potential formation mechanism may involve anomalously weak stellar feedback, due to low star formation rate surface densities. As a result of this, significant amounts of gas are not ejected from the disc, resulting in a galaxy which has a high baryon fraction and therefore a relatively low DM fraction \citep[e.g][]{Mancera_Pina2020}.

A further formation channel is DM stripping of satellite galaxies, particularly in extreme environments such as clusters \citep{Ogiya2018,Jing2019,Niemiec2019}. Indeed, both N-body simulations in a range of environments, from Milky-Way mass halos \citep{Hayashi2003,Kravtsov2004,Diemand2007,Rhee2017,Buck2019} to clusters \citep{Ghigna1998,Gao2004,Tormen2004,Nagai2005,Bosch2005,Giocoli2008,Xie2015}, and analytical models \citep[e.g.][]{Mamon2000,Gan2010,Han2016,Hiroshima2018} have shown that DM stripping is capable of removing parts of a galaxy's halo in group and cluster environments. It has been suggested that this process could drive a large scatter in the stellar-to-halo mass relation for satellite galaxies in groups and clusters, moving them from their initial positions towards lower halo masses \citep{Vale2004,Smith2016,Bah2017,Rhee2017,Niemiec2017,Niemiec2019}, and therefore lower M$_{\rm{halo}}$/M$_{\star}$ values. 

A comprehensive analysis of whether DM deficient systems (dwarfs in particular) can form naturally as a by-product of the process of galaxy evolution demands a hydrodynamical simulation in a cosmological volume which has both high mass and spatial resolution. The hydrodynamics is required for spatially-resolved predictions for the DM and baryons. The cosmological volume enables us to probe galaxy populations in a statistical manner, taking into account environmental effects (which idealised studies, for instance, cannot do), and is particularly important for making meaningful comparisons to large observational surveys \citep[e.g. forthcoming datasets like LSST, ][]{Robertson2019}. 

In recent years, large hydrodynamical cosmological simulations, e.g. Horizon-AGN \citep{Dubois2014}, Illustris \citep{Vogelsberger2014}, EAGLE \citep{Schaye2015} and Simba \citep{Dave2019}, have been successful in reproducing many properties of (massive) galaxies over cosmic time \citep[e.g.][]{Kaviraj2017}. For example, \citet{Saulder2020} have investigated the properties of massive (M$_{\star}$ > 10$^{9.5}$M$_{\odot}$) DM deficient galaxies which are isolated in these large cosmological simulations, akin to those found in \citet{Guo2020}. They find that these galaxies are probably regular objects that undergo un-physical processes at the boundary of the simulation box and are therefore artefacts. \citet{Jing2019} on the other hand have studied the formation of massive (10$^{9}$ M$_{\odot}$ < M$_{\star}$ < 10$^{10}$ M$_{\odot}$) DM deficient galaxies in the EAGLE and Illustris simulations, in denser environments. They find that a non-negligible fraction (2.6 per cent in EAGLE, and 1.5 per cent in Illustris) of satellite galaxies, in large groups and clusters (M$_{200}$ > 10$^{13}$ M$_{\odot}$), are DM deficient in their central regions. These galaxies, which are not initially DM deficient, become so through the stripping of DM by tidal interactions with their host galaxy, an effect that has also been noted in Horizon-AGN \citep{Volonteri2016}). 

However, given that these environments are relatively rare, and since the majority of dwarfs have lower stellar masses than the galaxies in these studies, it remains unclear from these studies (1) whether DM deficient galaxies can form in low-density environments which host the majority of objects, and (2) whether they can form in the dwarf regime where galaxies are expected to be significantly more DM dominated at early epochs, and where most observational studies that have found DM deficient galaxies are focussed. It is worth noting that it is challenging to probe dwarf galaxy evolution using the large-scale cosmological simulations mentioned above due to their relatively low mass and spatial resolutions. For example, the DM mass resolution is $\sim$10$^{8}$ M$_{\odot}$ in Horizon-AGN, EAGLE and Illustris and the spatial resolution of these simulations is around a kpc. Recall that the stellar scale height of the Milky Way, for example, is $\sim$300 pc \citep[e.g.][]{Kent1991,Corredoira2002,McMillan2011}, which implies that much better spatial resolution is needed to properly resolve dwarfs.   

An accurate exploration of the evolution of dwarf galaxies requires a cosmological simulation with significantly better mass and spatial resolution, in order to properly resolve the processes on the small spatial scales involved. In this study, we use the \texttt{NewHorizon} hydrodynamical cosmological simulation, to understand the formation of DM deficient galaxies in the stellar mass range M$_{\star}$ > 10$^{8}$ M$_{\odot}$. \texttt{NewHorizon} is a zoom-in of an average region within Horizon-AGN simulation, which has a volume of 142 comoving Mpc$^{3}$). \texttt{NewHorizon} offers a maximum spatial resolution of 34 pc and mass resolutions of 10$^4$ M$_{\odot}$ and 10$^6$ M$_{\odot}$ in the stars and DM respectively, making it an ideal tool to study the evolution of dwarf galaxies \citep[e.g.][]{Jackson2020,Martin2020}.
At the lower limit of our stellar mass range, a typical low-redshift dwarf galaxy contains $\sim$10,000 stellar and $\sim$10,000 DM particles respectively. Our aims are to (1) study whether DM deficient dwarf galaxies form naturally in \texttt{NewHorizon}, (2) estimate the frequency with which they are created and (3) quantify the processes that produce these galaxies. 

This paper is structured as follows. In Section \ref{sec:NH}, we briefly describe the \texttt{NewHorizon} simulation and the selection of DM deficient galaxies. In Section \ref{sec:formation}, we study the processes that create these objects. We summarise our findings in Section \ref{sec:summary}.


\section{Simulation}
\label{sec:NH}

The \texttt{NewHorizon} cosmological, hydrodynamical simulation \citep[][]{Dubois2020}, is a high-resolution simulation, produced using a zoom-in of a region of the Horizon-AGN simulation \citep[][H-AGN hereafter]{Dubois2014,Kaviraj2017}. H-AGN employs the adaptive mesh refinement code RAMSES \citep{Teyssier2002} and utilises a grid that simulates a 142 comoving Mpc$^{3}$ volume (achieving a spatial resolution of 1 kpc), using 1024$^3$ uniformly-distributed cubic cells that have a constant mass resolution, using \texttt{MPGrafic} \citep{Prunet2008}.

For \texttt{NewHorizon}, we resample this grid at higher resolution (using 4096$^3$ uniformly-distributed cubic cells), with the same cosmology as that used in H-AGN ($\Omega_m=0.272$, $\Omega_b=0.0455$, $\Omega_{\Lambda}=0.728$, H$_0=70.4$ km s$^{-1}$ Mpc$^{-1}$ and $n_s=0.967$; \citet{Komatsu2011}). The high-resolution volume occupied by \texttt{NewHorizon} is a sphere which has a radius of 10 comoving Mpc, centred on a region of average density within H-AGN. \texttt{NewHorizon} has a DM mass resolution of 10$^6$ M$_{\odot}$ (compared to 8$\times$10$^7$ M$_{\odot}$ in H-AGN), stellar mass resolution of 10$^4$ M$_{\odot}$ (compared to 2$\times$10$^6$ M$_{\odot}$ in H-AGN) and a maximum spatial resolution of 34 pc (compared to 1 kpc in H-AGN)\footnote{The gravitational force softening is equal to the local grid size.}. The simulation has been performed down to $z=0.25$. 


\subsection{Star formation and stellar feedback} 

\texttt{NewHorizon} employs gas cooling via primordial Hydrogen and Helium, which is gradually enriched by metals produced by stellar evolution \citep{Sutherland1993,Rosen1995}. An ambient UV background is switched on after the Universe is re-ionized at $z=10$ \citep{Haardt1996}. Star formation is assumed to take place in gas that has a hydrogen number density greater than $n_{H}>$10 H cm$^{-3}$ and a temperature lower than 2$\times$10$^4$ K, following a Schmidt relation \citep{schmidt1959,Kennicutt1998}. The star-formation efficiency is dependent on the local turbulent Mach number and the virial parameter $\alpha=2E_k$/|$E_g$|, where E$_k$ is the kinetic energy of the gas and E$_g$ is the gas gravitational binding energy \citep{Kimm2017}. The probability of forming a star particle of mass M$_{*,res}$=10$^4$ M$_{\odot}$ is drawn at each time step using a Poissonian sampling method, as described in \citet{Rasera2006}. 

Each star particle represents a coeval stellar population, with 31 per cent of the stellar mass of this population (corresponding to stars more massive than 6 M$_{\odot}$) assumed to explode as Type II
supernovae, 5 Myr after its birth. This fraction is calculated using a Chabrier initial mass function, with upper and lower mass limits of 150 M$_{\odot}$ and 0.1 M$_{\odot}$ \citep{Chabrier2005}. Supernova feedback takes the form of both energy and momentum, with the final radial momentum capturing the snowplough phase of the expansion \citep{Kimm2014}. The initial energy of each supernova is 10$^{51}$ erg and the supernova has a progenitor mass of 10 M$_{\odot}$. Pre-heating of the ambient gas by ultraviolet radiation from young O and B stars is included by augmenting the final radial momentum from supernovae following \citet{Geen2015}.


\subsection{Supermassive black holes and black-hole feedback}

Supermassive black holes (SMBHs) are modelled as sink particles. These accrete gas and impart feedback to their ambient medium via a fraction of the rest-mass energy of the accreted material. SMBHs are allowed to form in regions that have gas densities larger than the threshold of star formation, with a seed mass of 10$^4$ M$_{\odot}$. New SMBHs do not form at distances less than 50 kpc from other existing black holes. A dynamical gas drag force is applied to the SMBHs \citep{Ostriker1999} and two SMBHs are allowed to merge if the distance between them is smaller than 4 times the cell size, and if the kinetic energy of the binary is less than its binding energy.

The Bondi-Hoyle-Lyttleton accretion rate determines the accretion rate on to SMBHs, with a value capped at Eddington \citep{hoyle1939,Bondi1944}. The SMBHs release energy back into the ambient gas via a thermal quasar mode and a jet radio mode, when accretion rates are above and below 1 per cent of the Eddington rate respectively \citep{Dubois2012}. The spins of these SMBHs are evolved self-consistently through gas accretion in the quasar mode and coalescence of black hole binaries \citep{Dubois2014a}, which modifies the radiative efficiencies of the accretion flow (following the models of thin Shakura \& Sunyaev accretion discs) and the corresponding Eddington accretion rate, mass-energy conversion, and bolometric luminosity of the quasar mode \citep{Shakura1973}. The quasar mode imparts 15 per cent of the bolometric luminosity as thermal energy into the surrounding gas. The radio mode employs a spin-dependent variable efficiency and spin up and spin down rates that follow the simulations of magnetically-choked accretion discs \citep[see e.g.][]{Mckinney2012}. 


\subsection{Identification of galaxies and merger trees}

We use the AdaptaHOP algorithm to identify DM halos \citep{Aubert2004,Tweed2009}. AdaptaHOP efficiently removes subhalos from principal structures and keeps track of the fractional number of low-resolution DM particles within the virial radius of the halo in question. We identify galaxies in a similar fashion, using the HOP structure finder applied directly on star particles \citep{Eisenstein1998}. The difference between HOP and AdaptaHOP lies in the fact that HOP does not remove substructures from the main structure, since this would result in star-forming clumps being removed from galaxies. We then produce merger trees for each galaxy in the final snapshot at $z=0.25$, with an average timestep of $\sim$ 15 Myr, enabling us to track the main progenitor of every galaxy with high temporal resolution.

\begin{figure}
\centering
\includegraphics[width=\columnwidth]{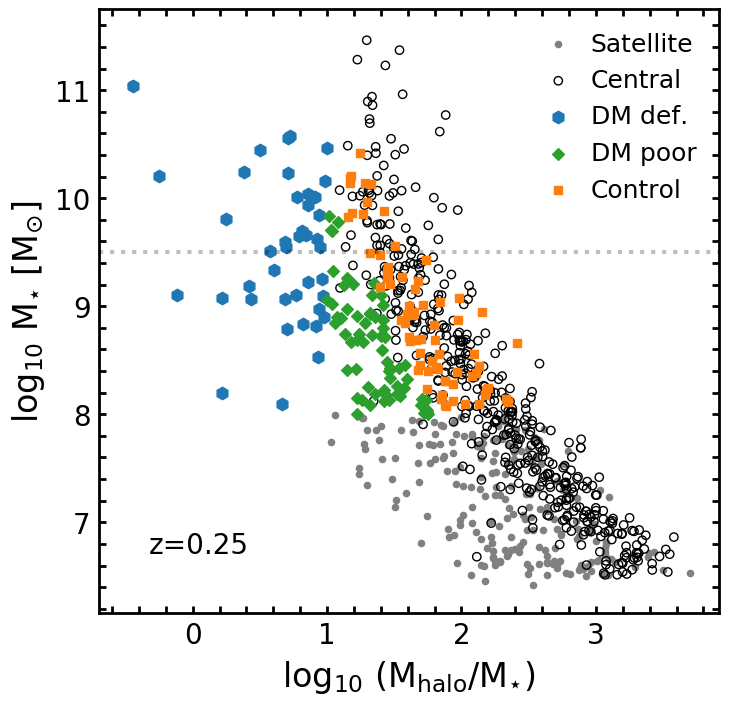}
\caption{The stellar mass (M$_{\star}$) vs the ratio of the DM halo (M$_{\rm{halo}}$) and stellar mass for galaxies in \texttt{NewHorizon} at $z=0.25$. In this study, we focus on galaxies with M$_{\star}$>10$^8$ M$_{\odot}$, which are well-resolved in the simulation to early epochs (galaxies that are less massive that this threshold are shown simply for completeness). We split this population into objects that are centrals (open black circles) and satellites (coloured circles). The satellites are further divided into `DM deficient' galaxies, defined as objects that exhibit M$_{\rm{halo}}$/M$_{\star}$<10 (blue circles), satellites that coincide with the relatively tight locus defined by the centrals (orange circles) and `DM poor' galaxies (green circles) which fall in between these two populations. The satellites that coincide with the centrals (orange circles) are essentially unstripped in their DM. We use these unstripped satellites as a control sample, since these galaxies show normal levels of DM. The grey dotted line denotes the mass threshold adopted for defining dwarf galaxies. While there are some galaxies in our sample above this threshold, the overwhelming majority of galaxies in our study are dwarfs.}
\label{fig:selection}
\end{figure}

Given that \texttt{NewHorizon} is a high resolution zoom of Horizon-AGN, it is worth considering the DM purity of galaxies, since higher-mass DM particles may enter the high resolution region of \texttt{NewHorizon} from the surrounding lower-resolution regions. Given the large mass difference, these DM particles may interact with galaxies they are passing through in unusual ways. The vast majority of galaxies affected by low DM purity exist at the outer edge of the \texttt{NewHorizon} sphere. The DM deficient galaxies studied in this paper all have DM halos with a purity of 100 per cent. 


\subsection{Selection of galaxies that are deficient in DM}
\label{select}

Figure \ref{fig:selection} shows the stellar mass (M$_{\star}$) vs the ratio of the DM halo (M$_{\rm{halo}}$) and stellar mass, for galaxies in \texttt{NewHorizon} at $z=0.25$. 
We study galaxies which have stellar masses above 10$^8$ M$_{\odot}$, which remain well-resolved in the simulation to early epochs. We note that, while there are some galaxies in our sample above the mass threshold that we use to define dwarfs (M$_{\star}$ < 10$^{9.5}$ M$_{\odot}$), the overwhelming majority of galaxies in our study are dwarfs. We split this galaxy population into objects that are centrals (open black circles) and satellites (coloured circles). Satellite galaxies are defined in the usual way, as systems whose DM halos have been accreted by a larger halo. The satellites are further divided into `DM deficient' galaxies (blue circles), defined as objects that exhibit M$_{\rm{halo}}$/M$_{\star}$<10, satellites that coincide with the relatively tight locus defined by the centrals (orange circles) and `DM poor' galaxies (green circles) which fall in between these populations. The satellites that occupy the same region in the M$_{\star}$-M$_{\rm{halo}}$/M$_{\star}$ plane as the centrals (orange circles) are essentially unstripped in their DM. For the analysis below, where we study the reasons for the DM stripping in the DM deficient and DM poor populations, we use these unstripped satellites as a control sample, since these galaxies show `normal' levels of DM (i.e. a similar DM content to that of centrals). 
\begin{figure}
\centering
\includegraphics[width=\columnwidth]{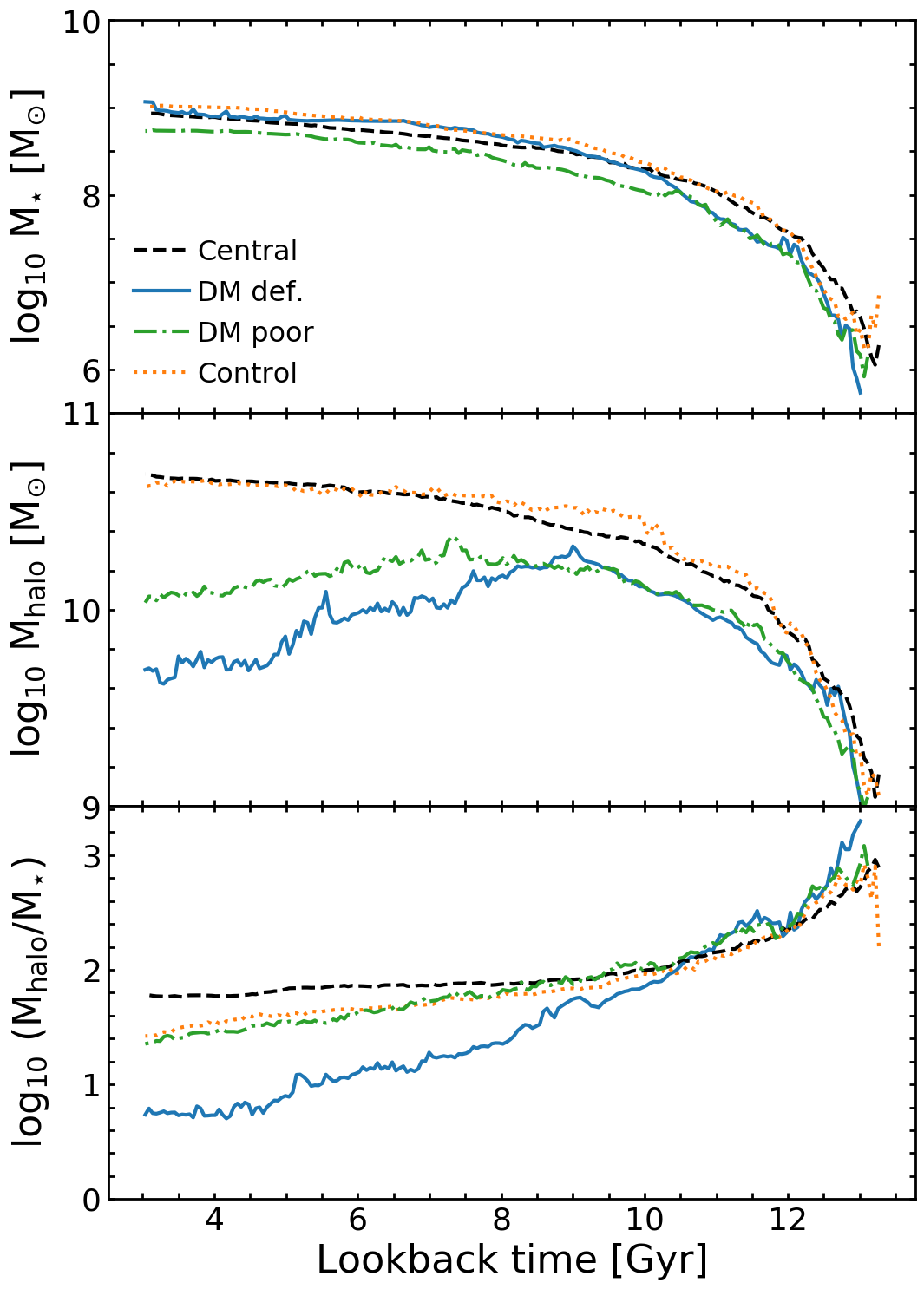}
\caption{The evolution of the median properties of the central, DM deficient, DM poor and control populations for M$_{\star}$>10$^8$ M$_{\odot}$. The top, middle and bottom panels show the evolution in the stellar mass, DM halo mass and the M$_{\rm{halo}}$/M$_{\star}$ ratio respectively. While the stellar mass evolution is similar across all populations, the DM halo mass evolves differently, with the DM poor and DM deficient galaxies exhibiting sharp declines in their halo mass at late epochs (i.e. look-back times less than 8 Gyrs). This, in turn, drives the scatter in the M$_{\rm{halo}}$ vs M$_{\star}$ relation seen in Figure \ref{fig:selection}.}
\label{fig:median_curves}
\end{figure}

\begin{figure*}
\centering
\includegraphics[width=\textwidth]{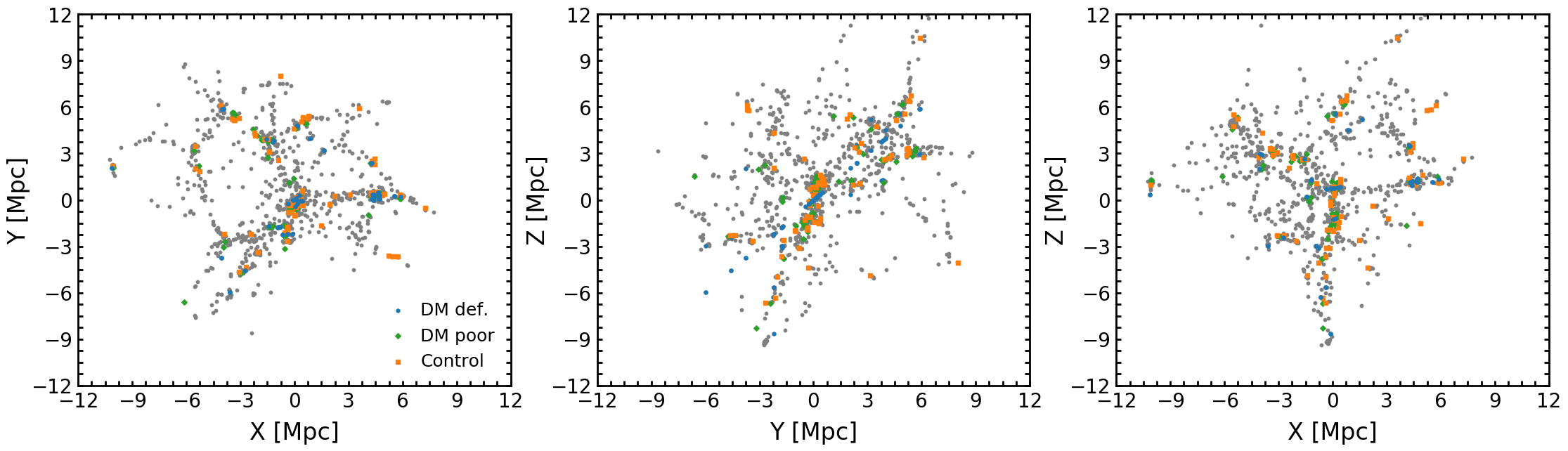}
\caption{The locations of our different galaxy populations (DM deficient, DM poor and control) in \texttt{NewHorizon} at $z=0.25$, in three mutually orthogonal projections (\textit{xy}, \textit{yz} and \textit{xz}) of the simulation volume. The centrals are shown using small grey circles. The DM deficient and DM poor populations form in all parts of the Universe, with the DM deficient galaxies generally showing a preference for the nodes of the cosmic web.}
\label{fig:locations}
\end{figure*}

\begin{figure}
\centering
\includegraphics[width=\columnwidth]{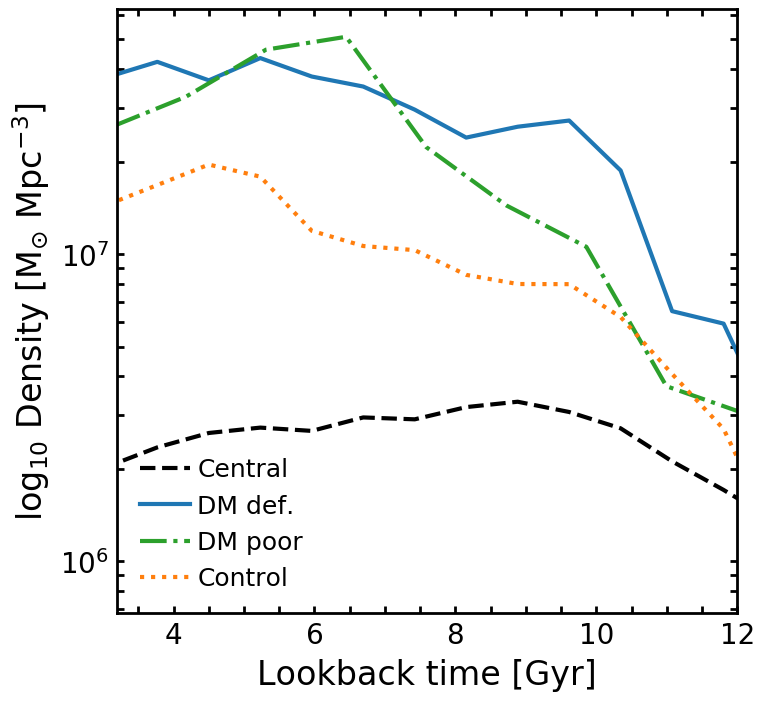}
\caption{The local 3D number density for the different populations vs look-back time for galaxies with M$_{\star}$>10$^8$ M$_{\odot}$. The local 3D number density is calculated using an adaptive kernel density estimation method (see text in Section \ref{sec:formation} for details). Recall that \texttt{NewHorizon} does not contain rich clusters and therefore the highest density regions in the simulation correspond to large groups, with halo masses of $\sim$10$^{13}$ M$_{\odot}$. Both the DM deficient and DM poor populations show elevated local densities, compared to the controls, indicating that local density may play a role in their formation.}
\label{fig:percentiles}
\end{figure}

Note that in this study we define M$_{\rm{halo}}$ to be the total mass of the DM halo associated with the galaxy. This is in contrast to previous studies \citep[e.g.][]{Jing2019} that have mostly looked at the central regions of the DM halo (e.g. two times the half-mass radius in the stars). The fractions of DM deficient, DM poor, control and central galaxies in \texttt{NewHorizon} are 12, 19, 18 and 51 per cent respectively. Recall that the centrals and the controls (i.e. the unstripped satellites) show normal levels of DM, so that the fraction of galaxies that show some sort of a deficiency in their DM content is around 30 per cent.

In Figure \ref{fig:median_curves} we show the evolution of the median properties of each population. The top, middle and bottom panels show the evolution in the stellar mass, DM halo mass and the M$_{\rm{halo}}$/M$_{\star}$ ratios respectively. While the stellar mass evolution is similar across all populations, the DM halo mass evolves differently, with the DM deficient and DM poor galaxies exhibiting declines in DM content at late epochs. The scatter in the M$_{\star}$ vs M$_{\rm{halo}}$/M$_{\star}$ relation seen in Figure \ref{fig:selection} is therefore driven by the removal of DM in these galaxies, which produces systems that are poorer in DM for their stellar mass.

\begin{figure*}
\centering
\includegraphics[width=\columnwidth]{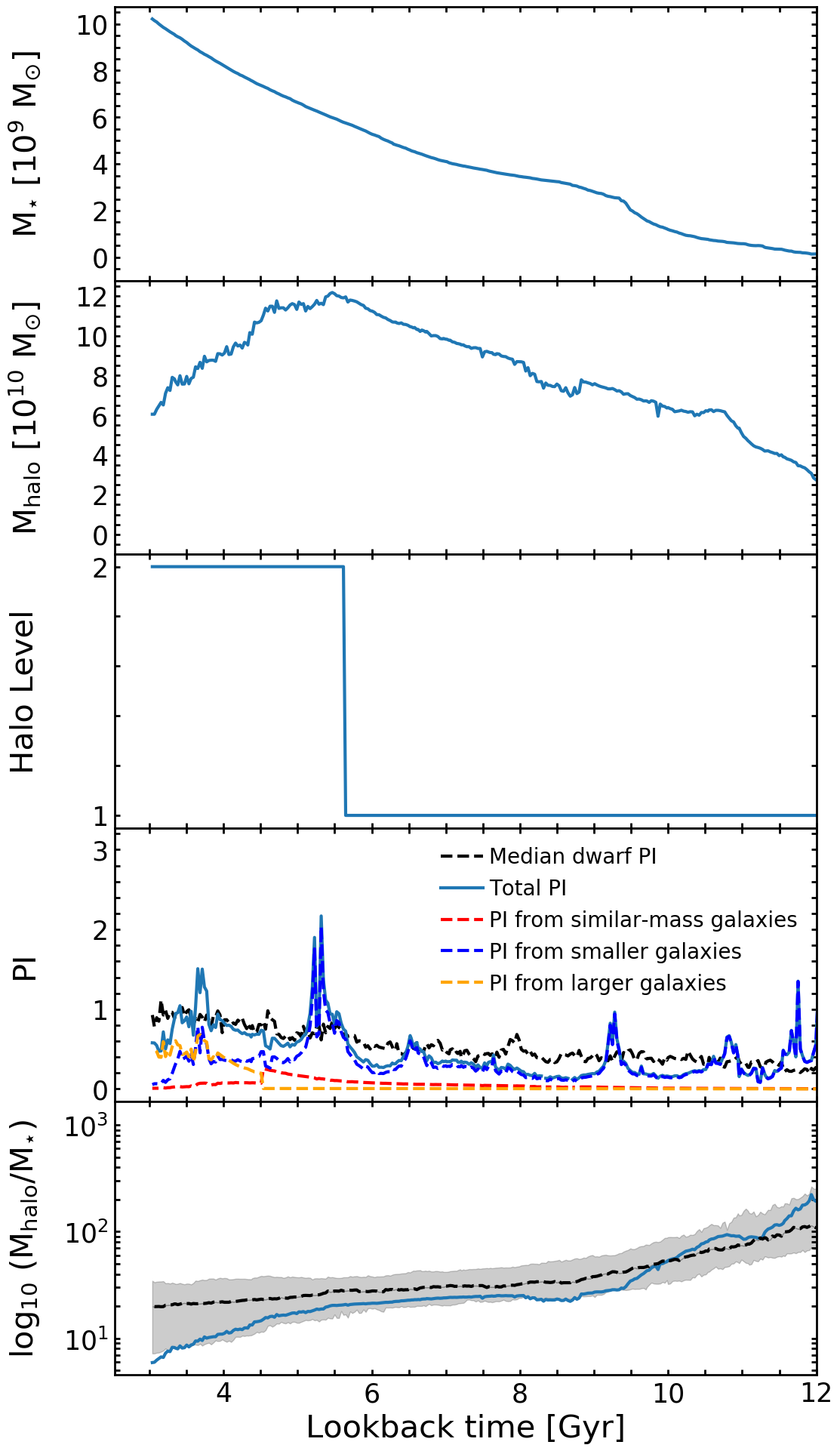}
\includegraphics[width=\columnwidth]{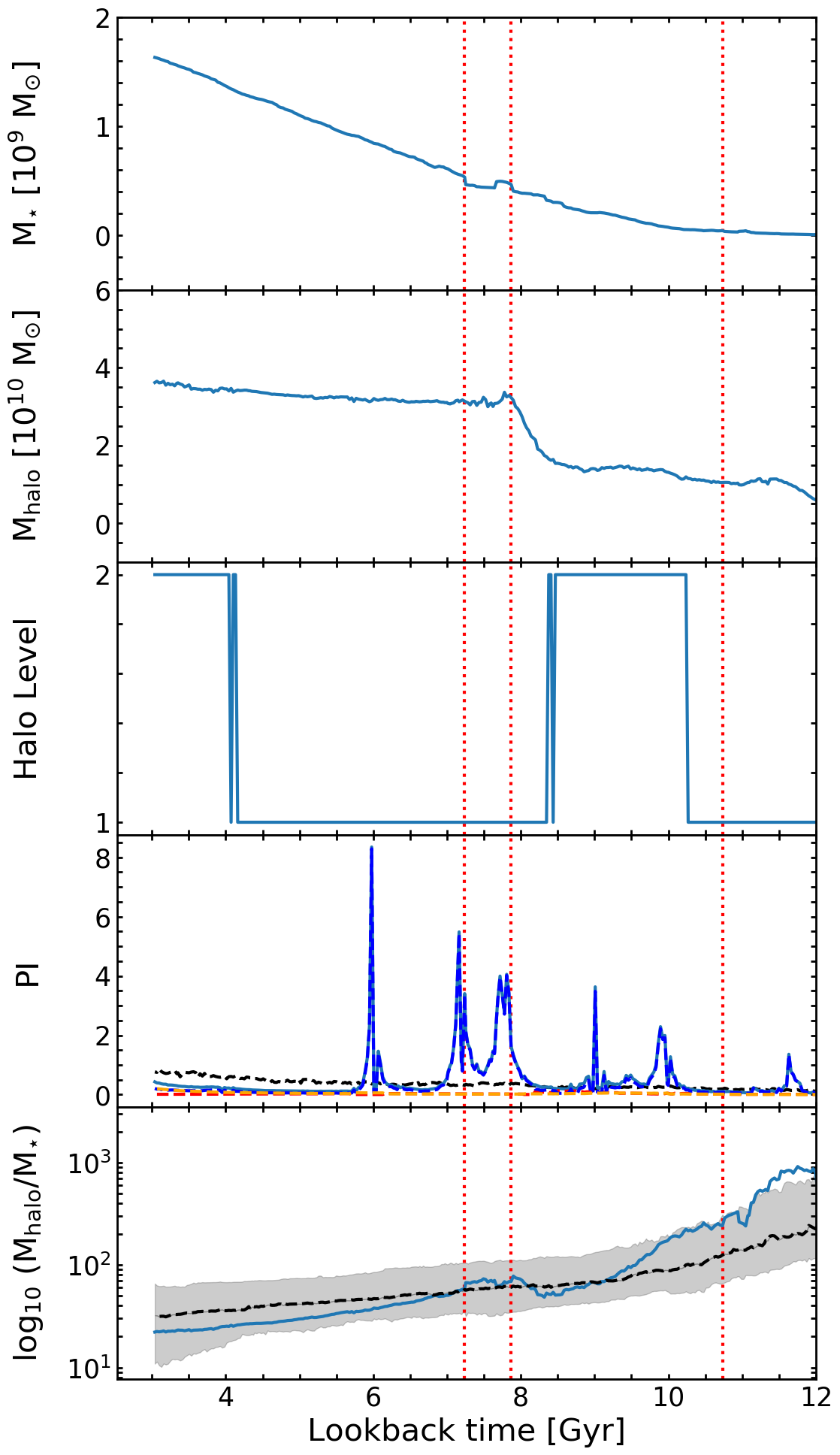}
\caption{The rows show (from top to bottom) the evolution of the stellar mass, DM halo mass, the halo level (central [level 1] or satellite [level > 1]), perturbation index (PI) and the M$_{\rm{halo}}$/M$_{\star}$ ratio of the galaxy in question. The PI is split into contributions from `similar-mass' galaxies (those that have 0.3 M$_{\rm{gal}}$ < M$_{\star}$ < 3 M$_{\rm{gal}}$, where M$_{\rm{gal}}$ is the stellar mass of the galaxy in question), `smaller galaxies' (those that have M$_{\star}$ < 0.3 M$_{\rm{gal}}$) and `larger galaxies' (those that have M$_{\star}$ > 3 M$_{\rm{gal}}$). The `median dwarf PI' indicates the median PI experienced by galaxies with stellar masses within 0.5 dex of the galaxy in question. In the bottom panel, the dashed black line indicates the median halo-to-stellar mass ratio of galaxies with stellar masses within 0.5 dex of the galaxy in question and the shaded region is its 1$\sigma$ dispersion. The left-hand column shows a DM deficient galaxy, while the right-hand column shows a control galaxy. The dotted red lines indicate times when galaxy mergers, with mass ratios greater than 10:1, take place.}
\label{fig:dmevo}
\end{figure*}

While the gradual change from the locus of centrals and controls seen in Figure \ref{fig:selection} indicates that the DM deficient and DM poor populations are not simply artefacts and are likely created by a persistent process that operates on these galaxies, we perform several checks to ensure that this is indeed the case. The galaxies in these populations are found in all parts of the simulation sphere (as discussed further in Section \ref{sec:formation} and Figure \ref{fig:locations}) and not just at the edges of this volume where boundary effects can produce spurious effects. They also exhibit 100 per cent purity in DM, so that their evolution is not driven by interactions with massive low-resolution DM particles passing through them. Finally, as shown in the analysis below (Figure \ref{fig:dmevo}) the reduction of DM is very gradual, rather than abrupt, which indicates that this phenomenon is not a result of misallocation of particles between DM halos. 



\section{Formation of DM deficient galaxies through stripping of DM in tidal interactions}
\label{sec:formation}

We begin our analysis by exploring the local environment of our different satellite populations (DM deficient, DM poor, and control). Figure \ref{fig:locations} shows the positions of these populations in the cosmic web. The DM deficient and DM poor populations form in all parts of the Universe, with the DM deficient galaxies generally showing a preference for the nodes of the cosmic web. To explore local galaxy density more quantitatively, we follow \citet{Martin2018a} and define a 3-D local number density of objects around each galaxy. The local density is calculated using an adaptive kernel density estimation method\footnote{Code available at \href{https://github.com/garrethmartin/MBEtree}{github.com/garrethmartin/MBEtree}}, where the width of the kernel used for multivariate density estimation is responsive to the local density of the region, such that the error between the density estimate and the true density is minimised \citep{Breiman1977,Ferdosi2011,Martin2018a}. The density estimate takes into account all galaxies with stellar masses above 10$^{8}$ M$_{\odot}$. In Figure \ref{fig:percentiles} we show the evolution of the local density occupied by our different populations. Recall that \texttt{NewHorizon} does not contain rich clusters and therefore the highest density regions in the simulation correspond to large groups with halo masses of $\sim$10$^{13}$ M$_{\odot}$.


\begin{figure}
\centering
\includegraphics[width=\columnwidth]{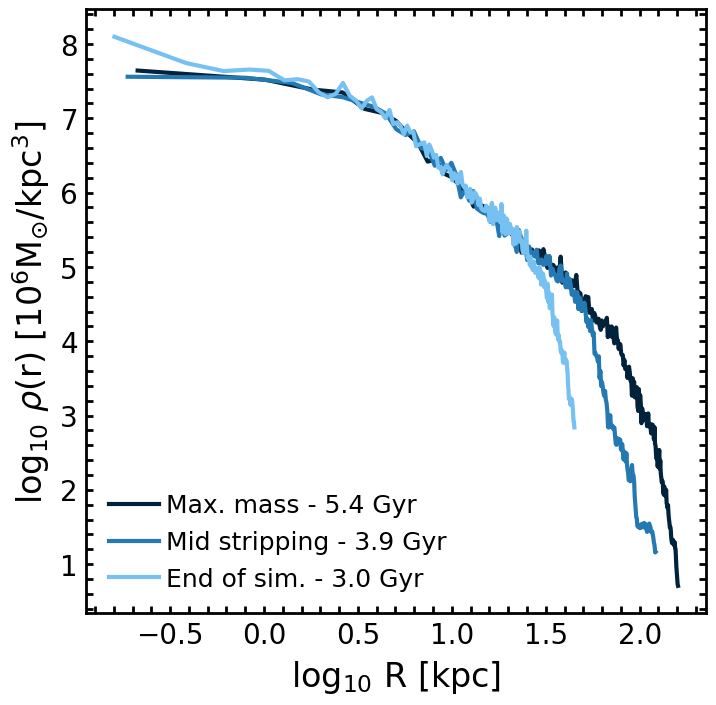}
\caption{Evolution of the DM density profile, for various lookback times, in the DM deficient galaxy described in the left-hand column of Figure \ref{fig:dmevo}. The evolution is shown from the point at which the galaxy becomes a satellite, which is also the point at which the DM stripping starts (`max mass'), through to the end of the simulation. The DM stripping takes place in the outskirts of the dwarf where the gravitational potential well is shallowest and DM particles are more loosely bound.}
\label{fig:profiles}
\end{figure}

\begin{figure}
\centering
\includegraphics[width=\columnwidth]{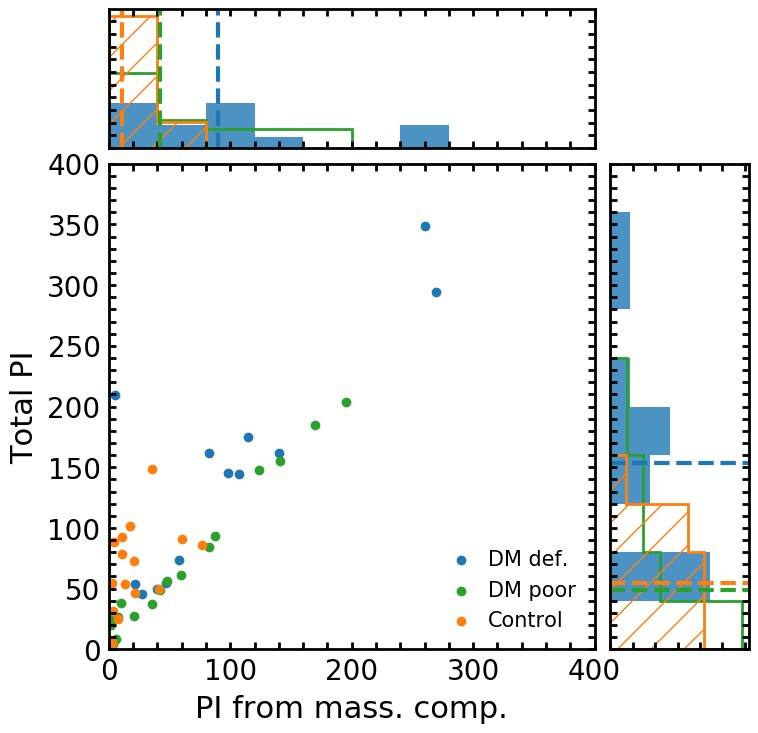}
\caption{The total perturbation index (PI) from all companions vs that just from the massive companion, integrated over the period from when the dwarf becomes a satellite until the end of the simulation. The different satellite populations are shown using different colours. The histograms are normalised and the dashed lines indicate median values. DM deficient galaxies show both larger values for the total PI and the PI from massive companions, compared to their control counterparts (with the DM poor galaxies falling in between these two populations). Note also that the PI from the massive companion dominates the total PI in the DM deficient and DM poor populations.}
\label{fig:totlarge}
\end{figure}

\begin{figure*}
\centering
\includegraphics[width=\textwidth]{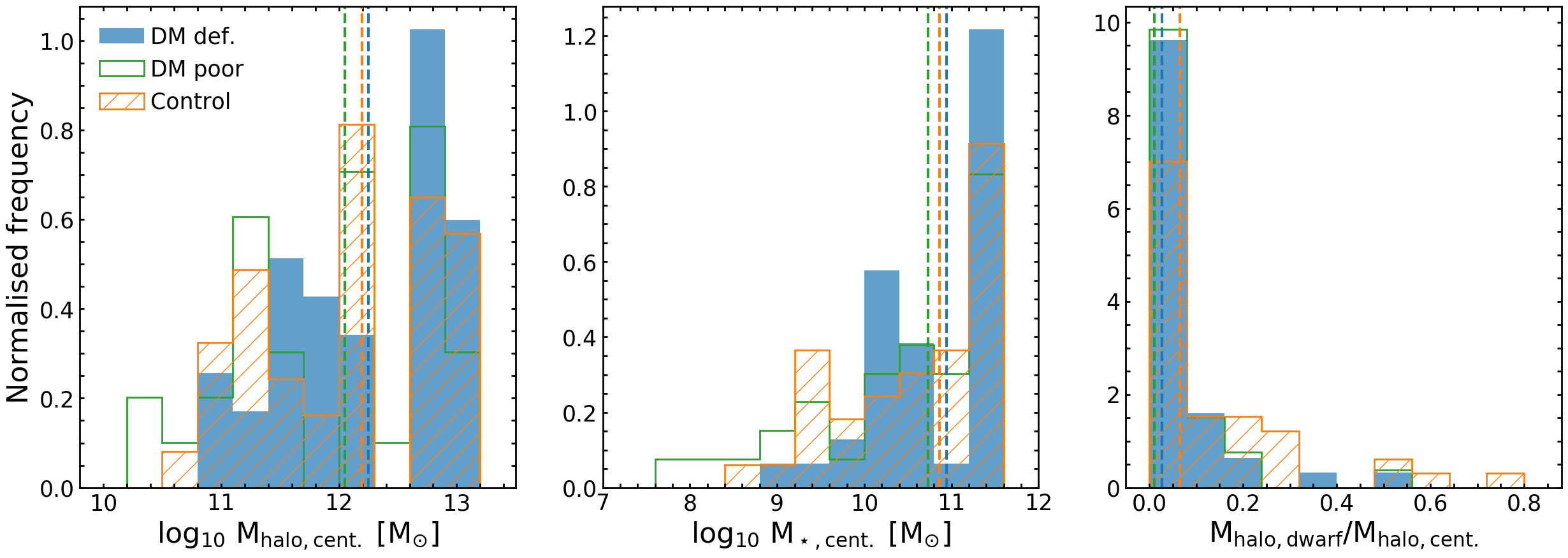}
\caption{The DM halo (left) and stellar (centre) mass of the massive centrals that host our dwarf satellites and the halo mass ratio (right) between the dwarf satellite and its massive central. The dashed lines indicate the median values of the histograms. The distribution of massive central masses and halo mass ratios are similar for all dwarf populations and are therefore not the principal reason for the differences in perturbation index (PI) in these populations.}
\label{fig:host_mass}
\end{figure*}

\begin{figure*}
\centering
\includegraphics[width=\textwidth]{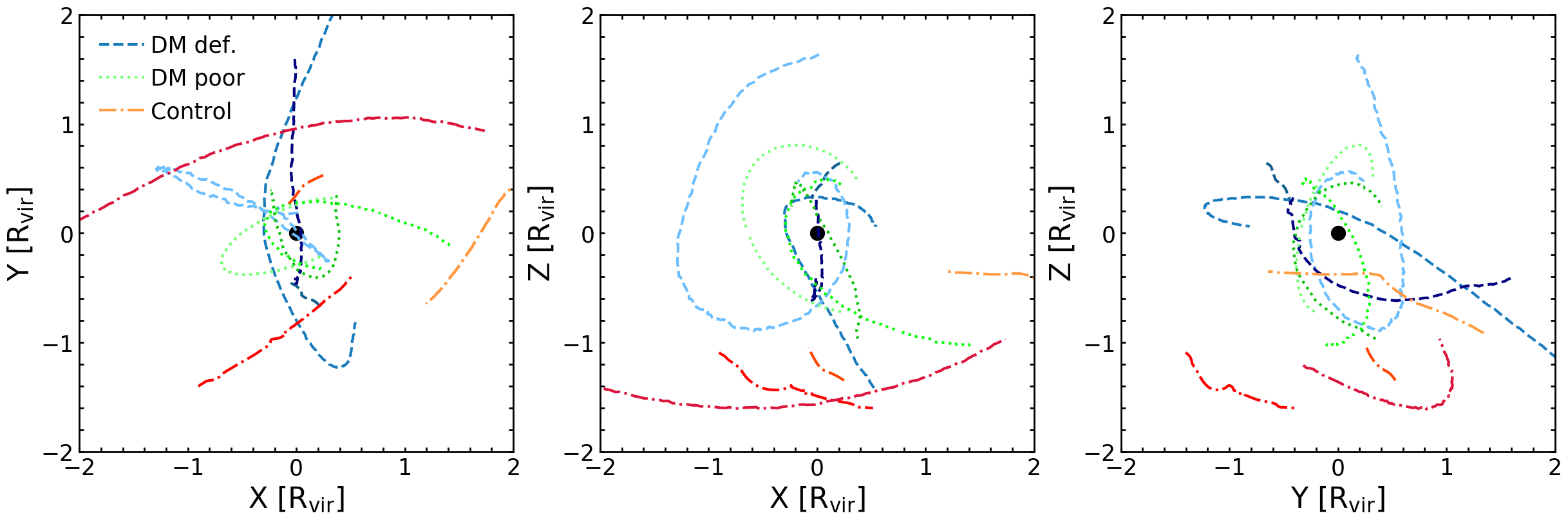}
\caption{The orbits of four galaxies in each of the DM deficient (shades of blue), DM poor (shades of green) and control (shades of orange) populations around their centrals. Individual galaxy orbits are shown using different shades of the colour in question (see legend). We show orbits from the point at which the dwarf in question becomes a satellite through to the end of the simulation. The orbital distances are normalised by the virial radius of the central, which is always shown at the origin.}
\label{fig:orbits}
\end{figure*}

\begin{figure*}
\centering
\includegraphics[width=\textwidth]{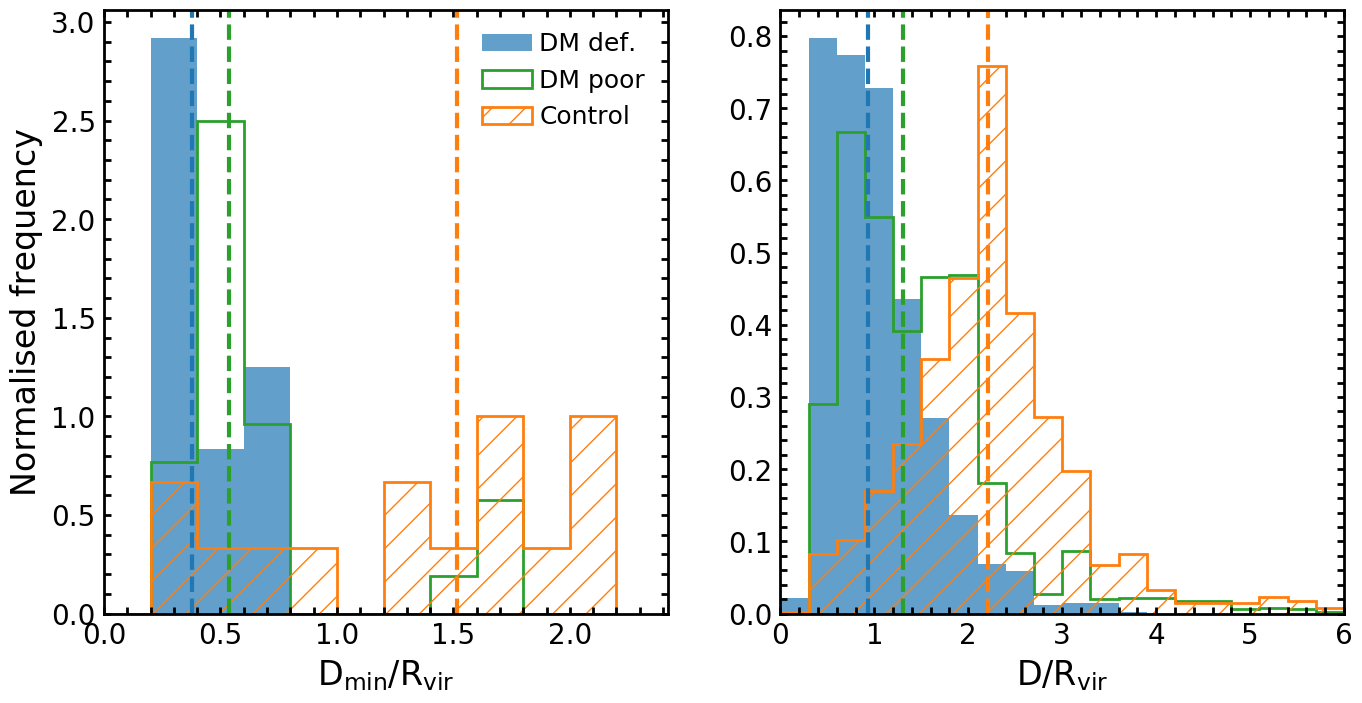}
\caption{\textbf{Left:} Minimum distance between dwarfs and their massive centrals for our different satellite populations. The dashed lines indicate the median value of each distribution. \textbf{Right:} Orbital distances (normalised by the virial radius of the central) between the dwarf and their massive central at all timesteps in the period between the dwarf becoming a satellite and the end of the simulation. The DM deficient and DM poor populations exhibit closer orbits than their control counterparts which drives the higher perturbation index (PI) values seen in these populations.}
\label{fig:alldist}
\end{figure*}

\begin{figure*}
\centering
\includegraphics[width=\textwidth]{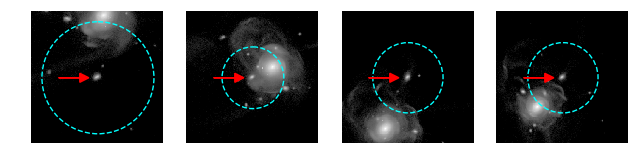}
\includegraphics[width=\textwidth]{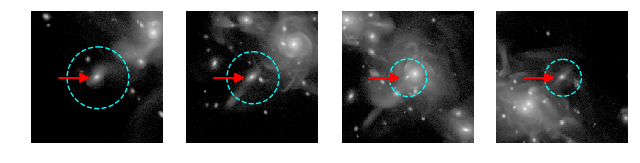}
\caption{$r$-band mock images of two example systems that bracket the types of interactions that lead to DM stripping. Each image is 200 proper kpc across. In all images, the DM deficient galaxy in question is at the centre of the image and indicated by the red arrow. The blue circle shows the virial radius of the DM deficient galaxy. The sequence (where time moves forward from left to right) shows $\sim$3 Gyrs of evolution. The top row shows a two body interaction, while the bottom row shows a more complex interaction where the dwarf interacts with a larger companion which dominates a group. In both cases, the dwarf undergoes significant DM stripping and ends up as a DM deficient galaxy.}
\label{fig:images_4}
\end{figure*}



Figures \ref{fig:locations} and \ref{fig:percentiles} indicate that the DM deficient and DM poor populations form in all regions of the Universe (note again that \texttt{NewHorizon} does not contain clusters or large voids so these environments are not probed here). The DM deficient galaxies generally show a preference for the nodes of the cosmic web. Both the DM deficient and DM poor populations show elevated local densities, compared to the controls, indicating that local density may play a role in their formation. Not unexpectedly, the dwarf centrals reside in lower density environments, where the probability of interacting with, and becoming the satellite of, a more massive galaxy is lower. 

The evolution of galaxies is influenced by a combination of internal processes, like supernova \citep[e.g.][]{Kimm2014} and AGN \citep[e.g.][]{Croton2006} feedback, and external processes such as tidal perturbations \citep[e.g.][]{Martin2019}, mergers \citep[e.g.][]{Kaviraj2014a,Kaviraj2014b,Kaviraj2019} and ram pressure \citep[e.g.][]{Hester2006}. Baryonic feedback and ram pressure are not capable of removing significant amounts of DM in galaxies, except in the very central regions through rapid expansion of gas \citep{Navarro1996,Governato2012,Teyssier2013}. Furthermore, the steady stripping of DM, which drives the creation of the DM poor and DM deficient populations, requires a dynamical process, suggesting that tidal perturbations (possibly including mergers) are likely to be important in giving rise to these systems. In particular, interactions between a massive galaxy and a lower mass companion typically results in the stripping of material from the smaller companion. This is because, while the same tidal force acts on both objects, the smaller companion is not as strongly bound and stripping typically takes place from the outskirts of the system where the depth of the gravitational potential well is shallowest \citep{Smith2016}. 

To quantify the effect of tidal perturbations, we follow \citet{Martin2019} and \citet{Jackson2020} to define a dimensionless `perturbation index' (PI) that quantifies the strength of the ambient tidal field around a galaxy:

\begin{equation}
\label{eqn:PI}
\rm{PI = \sum_{i}\left ( \frac{M_{i}}{M_{halo}}\right ) \left ( \frac{R_{\mathrm{vir}}}{D_{i}} \right )^{3}},
\end{equation}

\noindent where M$_{\rm{halo}}$ is the halo mass of the galaxy in question, R$_{\mathrm{vir}}$ is the virial radius of its DM halo, M$_{\rm{i}}$ is the DM halo mass of the $i$th perturbing galaxy and D$_{\rm{i}}$ is the distance to the $i$th perturbing galaxy. We consider all perturbing galaxies within 3 Mpc of the object in question. 

We use the PI to explore the trends in the tidal perturbations in our different populations and study how these perturbations may be driving the DM stripping in the DM poor and DM deficient populations. In Figure \ref{fig:dmevo}, we visually illustrate the evolution of two typical objects (selected randomly) in the DM deficient and control populations. The left-hand column shows the DM deficient galaxy, while the right-hand column shows the control galaxy. Recall that the control galaxies are satellites in which DM is not stripped and that the DM deficient galaxies are systems which have the lowest values of M$_{\rm{halo}}$/M$_{\star}$ (Figure \ref{fig:selection}). These two populations therefore bracket the dwarf population as a whole. The rows show (from top to bottom) the evolution of the stellar mass, DM halo mass, halo level (central [level 1] or satellite [level > 1]), PI and the M$_{\rm{halo}}$/M$_{\star}$ ratio of the galaxy in question. 

The PI is split into contributions from `similar-mass' galaxies (those that have 0.3 M$_{\rm{gal}}$ < M$_{\star}$ < 3 M$_{\rm{gal}}$, where M$_{\rm{gal}}$ is the stellar mass of the galaxy in question), `smaller galaxies' (those that have M$_{\star}$ < 0.3 M$_{\rm{gal}}$) and `larger galaxies' (those that have M$_{\star}$ > 3 M$_{\rm{gal}}$). While the DM deficient galaxy shows a steady increase in stellar mass (row 1), it exhibits sustained DM stripping after a look-back time of around 6 Gyrs (row 2). The stripping coincides with the galaxy transitioning from being a central to a satellite (row 3) and an increase in the PI from massive galaxies, even as that from smaller galaxies decreases (row 4). The DM stripping drives the M$_{\rm{halo}}$/M$_{\star}$ ratio down, until it is below the region traced by the median ratio for galaxies which have stellar masses within $\pm$0.5 dex (row 5). The principal difference between the DM deficient galaxy and its control counterpart is that, even after the control object becomes a satellite, the PI due to massive galaxies does not increase and the control galaxy does not exhibit significant DM stripping.

In Figure \ref{fig:profiles}, we show the evolution of the DM density profile (calculated using all DM particles associated with the galaxy) of the DM deficient galaxy described in the left-hand column of Figure \ref{fig:dmevo}. The evolution is shown from the point at which the galaxy becomes a satellite (which is also the point at which the DM stripping starts), through to the end of the simulation. Not unexpectedly, the DM stripping typically takes place in the outskirts of the dwarf, where the gravitational potential well is shallower and DM particles are less well bound. The trends are identical in all DM deficient and DM poor galaxies. 

\begin{table*}
\begin{tabular}{| c | c | c | c |}
\hline
\hline
PI & PI fraction & Cumulative PI & Mass ratio \\
\hline
\hline
0.48 & 0.78 & 0.78 & 5.75 \\
0.05 & 0.08 & 0.86 & 0.01 \\
0.01 & 0.02 & 0.88 & 0.88 \\
0.01 & 0.01 & 0.89 & 0.01 \\
\end{tabular}
\begin{tabular}{| c | c | c | c |}
\hline
\hline
PI & PI fraction & Cumulative PI & Mass ratio \\
\hline
\hline
0.09 & 0.25 & 0.25 & 10.98 \\
0.04 & 0.12 & 0.37 & 0.67  \\
0.04 & 0.12 & 0.49 & 0.02  \\
0.02 & 0.04 & 0.53 & 0.01  \\
\end{tabular}
\caption{The four largest perturbation index (PI) contributions from individual companions for the DM deficient (left hand table) and control (right hand table) galaxy shown in Figure \ref{fig:dmevo}, at the point where the galaxy becomes a satellite. Columns in each table show (from left to right) the PI contributed by the individual companion, the fraction this represents of the total PI, the cumulative fraction of PI contributed by this companion and those in previous rows and the mass ratio of the companion and the dwarf galaxy in question. The left hand table, which corresponds to the DM deficient galaxy, shows that $\sim$80 per cent of the total PI comes from one larger companion with the next largest PI contribution being an order of magnitude smaller. For the control galaxy the largest PI contribution, while also derived from a larger companion, is a factor of 5 smaller than that in the DM deficient galaxy and contributes only 25 per cent of the total PI, with many smaller companions contributing PI values which are within a factor of 2 of that from the larger companion.}
\label{tab:PI values}
\end{table*}

In Table \ref{tab:PI values}, we present the four largest PI contributions from individual companions, for the DM deficient and control galaxy in Figure \ref{fig:dmevo}, at the point where each galaxy becomes a satellite. The left hand table, which corresponds to the DM deficient galaxy, shows that $\sim$80 per cent of the total PI comes from one larger companion (which is the central that hosts this dwarf), with the next largest PI contribution being an order of magnitude smaller. For the control galaxy, on the other hand, the largest PI contribution, while also derived from a larger companion (again the host central), is a factor of 5 smaller than that in the DM deficient galaxy and contributes only 25 per cent of the total PI. Indeed, many smaller companions contribute PI values which are within a factor of 2 of that from the larger companion. 
Note that, in all cases, the larger companion which dominates the PI in the DM deficient and DM poor galaxies is the central that hosts the dwarf satellite in question. While Figure \ref{fig:dmevo} illustrates the trends seen in typical galaxies in the DM deficient and control populations, Figure \ref{fig:totlarge} summarises the differences in the PI from massive galaxies seen in our different satellite populations, after the galaxies become satellites. We show both the total PI from all companions vs that just from the massive companion, integrated over the period from when the dwarf becomes a satellite until the end of the simulation. DM deficient galaxies show both larger values for the total PI and the PI from massive companions, compared to their control counterparts (with the DM poor galaxies falling in between these two populations). Note also that the PI from the massive companion dominates the total PI in the DM deficient and DM poor populations. 
This indicates that the DM stripping in the DM deficient and DM poor galaxies is indeed driven by stronger interactions with a massive (central) companion. 

We proceed by exploring why the DM deficient dwarfs exhibit higher values of PI due to their massive companions. Figure \ref{fig:host_mass} indicates that both the halo (left panel) and stellar (centre panel) masses of the central that host the dwarfs are similar. In addition, the mass ratios between the dwarfs and their host centrals (right panel), are comparable across all the different populations. Therefore, these quantities are unlikely to be driving the higher values of PI in the DM deficient and DM poor populations. 

However, given the strong dependence of the PI on the distance between galaxies (Equation 1), it is likely that the higher PI in DM deficient and DM poor galaxies is driven by orbits that bring them closer to their corresponding centrals. In Figure \ref{fig:orbits}, we first illustrate this graphically by plotting the orbits of four galaxies (selected to span the mass range in our study) in each of the DM deficient, DM poor and control populations around their centrals. The orbital distances are normalised by the virial radius of the central and orbits are shown from the point at which the dwarf in question becomes a satellite to the end of the simulation. It is clear that DM deficient galaxies have orbits that bring them significantly closer to their massive central than in their control counterparts. 

Figure \ref{fig:alldist} presents this more quantitatively, by comparing both the orbital distances (normalised by the virial radius of the central) in the period after individual galaxies become satellites, and the minimum distances between the satellites and centrals during these orbits. DM deficient dwarfs spend most of their orbits at significantly smaller distances from their centrals compared to their control counterparts, with the DM poor galaxies lying in between these two populations. The same patterns are apparent in the minimum orbital distances. The larger PI values from larger galaxies in these populations are therefore driven by orbits which bring these galaxies closer to their massive central companions. The degree of DM stripping is correlated with these orbital distances, with members of the DM deficient population (which exhibits the largest amount of stripping) spending their orbits at much smaller distances than their DM poor and control counterparts. 

Given that the DM deficient galaxies are in tight orbits around nearby massive companions, it is worth considering how long these dwarfs may survive. We explore this by considering the evolution of the DM deficient population (selected in an identical way) at $z=0.7$ i.e. around 3.5 Gyrs before the epoch ($z=0.25$) at which the analysis above is performed. The stellar mass distribution of DM deficient galaxies at both epochs is similar. We find that only 30 per cent of DM deficient galaxies that exist at $z=0.7$ still survive at $z=0.25$. This indicates that the creation of DM deficient galaxies is a constant process over cosmic time. In other words, satellites that are in close orbits are not only stripped of their DM, but many also do not survive after the stripping starts as they are accreted by their larger companions. Thus, the DM deficient population at a given redshift are largely systems that have formed recently enough that they still exist in the simulation at that epoch. 

We also consider whether dwarfs that may be candidates for being DM deficient systems could potentially be identified using quantities that are readily available in imaging survey data. As described above, the process that drives the creation of these objects are tidal interactions with more massive companions on very close orbits. In the appendix we show a version of the right-hand panel of Figure \ref{fig:alldist} without normalising the orbital distances of the dwarfs by the virial radius of the massive companion (Figure \ref{fig:dist_nonorm}), because virial radii are not measurable quantities in imaging data. Combining Figures \ref{fig:host_mass}, \ref{fig:alldist}, \ref{fig:images_4} and \ref{fig:dist_nonorm} indicates that dwarfs that are found close to massive companions with M$_{\star}$ > 10$^{10}$ M$_{\odot}$ (Figure \ref{fig:host_mass}) at distances less $\sim$150 kpc (Figure \ref{fig:dist_nonorm}) and show stellar tidal features (Figure \ref{fig:images_4}) are likely to be good candidates for being DM deficient systems. As is typical of interactions between massive galaxies and dwarf companions \citep[e.g.][]{Kaviraj2010,Kaviraj2014b}, the tidal features (e.g. the ones visible in Figure \ref{fig:images_4}) are faint with surface brightnesses typically fainter than 29 mag arcsec$^{-2}$. This suggests that finding large samples of DM deficient candidates would ideally require deep wide-area surveys, such as LSST \citep{Robertson2019,Kaviraj2020}, which have limiting surface-brightnesses that are fainter than such values. 

\subsection{Comparison to observational studies}

We complete our study by considering our results in the context of recent observational work. DM deficient galaxies form in \texttt{NewHorizon} via tidal interactions in close orbits with a more massive (central) companion. It is worth noting that the theoretical study of \citet{Maccio2020} comes to similar conclusions, albeit using a small sample of three zoom-in simulations of individual galaxies. The formation mechanism described in our study is strongly supported by the recent empirical work of \citet{Montes2020}, who have used deep optical imaging to demonstrate that the DM deficient galaxy NGC 1052-DF4 exhibits tidal tails, similar to those predicted in Figure \ref{fig:images_4}. They show that these tidal features are driven by an interaction with its neighboring galaxy NGC 1035, and argue that the DM-deficient nature of this dwarf galaxy is caused by the DM stripping produced by this interaction.

Our proposed mechanism is also consistent with the recent studies by \citet{vanDokkum2018} and \citet{vanDokkum2019}, who find multiple DM deficient dwarf galaxies in group environments. These galaxies are likely to have been subjected to strong tidal forces that would have stripped their DM halo (see e.g. the lower panel of Figure \ref{fig:images_4} which shows a possible analog of this in \texttt{NewHorizon}). The formation of DM deficient dwarfs via tidal interactions also appears consistent with the excess number of globular clusters (GCs) reported in these systems \citep[e.g.][]{Fensch2019,Muller2020}. Since tidal interactions between dwarfs and larger companions can trigger enhanced GC formation \citep[e.g.][]{Fensch2019b,Carleton2020,Somalwar2020}, systems like DM deficient dwarfs, that undergo strong tidal interactions, could be expected to show an excess number of GCs. Note, however, that the resolution of \texttt{NewHorizon} is not sufficient for us to directly study the formation of GCs. 


The formation channel outlined in this paper appears less well-aligned with the findings of \citet{Guo2020}, who suggest that most of their DM deficient galaxies lie at distances greater than three times the virial radius of the nearest group or cluster. However, it is worth noting that the nearest groups and clusters in Guo et al. are significantly more massive than those in our study. It is possible that the tidal forces required to strip dwarfs of their DM can be produced at larger distances around much more massive groups. Furthermore, without deep images it is difficult to identify tidal features which are the tell-tale signatures of the tidal stripping process that creates DM deficient galaxies. Nevertheless, the relatively isolated nature of these galaxies, coupled with other potential mechanisms for producing such objects, e.g. weak baryonic feedback \citep{Mancera_Pina2020} suggests that other channels maybe required to satisfactorily explain the entire DM deficient population.

\section{Summary}
\label{sec:summary}
In the standard $\Lambda$CDM paradigm, dwarf galaxies are expected to be DM-rich, because their shallow gravitational potential wells make it easier for processes like stellar and supernova feedback to deplete their gas reservoirs. This results in a rapid depletion of gas and reduction of star formation at early epochs, leaving these objects with relatively high DM fractions. However, recent observational work suggests that some local dwarfs exhibit DM fractions as low as unity, around 400 times lower than what is expected for systems of their stellar mass. The existence of such DM deficient galaxies appears to contradict our classical expectations of the DM properties of dwarf galaxies, potentially bringing the validity of the standard paradigm into serious question. Understanding the origins of these galaxies, using a high-resolution cosmological simulation which can make realistic statistical predictions of dwarf galaxies, is, therefore, a key exercise. 

Here, we have used the \texttt{NewHorizon} cosmological hydrodynamical simulation to explain the formation of DM deficient galaxies. We are able to perform this exercise, for the first time, in a statistical fashion, as the cosmological volume of  \texttt{NewHorizon} allows us to study large numbers of dwarf galaxies, while its high spatial resolution enables us to resolve these systems with the requisite detail. We have shown that interactions between massive central galaxies and dwarf satellites can drive sustained and significant stripping of DM from the dwarfs, which reduces their DM content. The level of stripping is determined by the details of the orbit, with dwarfs that are heavily stripped typically spending significant fractions of their time in tight orbits around their corresponding massive central, after they turn into satellites. 

DM stripping is responsible for the large dispersion in the stellar-to-halo mass relation in the dwarf regime (see Figure \ref{fig:selection}), with the DM deficient and DM poor populations, which comprise $\sim$30 per cent of dwarfs, scattering off the tight main locus defined by the dwarf centrals and the dwarf satellites that remain unstripped. In extreme cases, this DM stripping produces dwarfs which exhibit M$_{\rm{halo}}$/M$_{\star}$ ratios as low as unity, consistent with the findings of recent observational studies. Given their close orbits, a significant fraction of DM deficient dwarfs will merge with their massive companions and disappear from the galaxy population (e.g. $\sim$70 per cent of such dwarfs will merge over timescales of $\sim$3.5 Gyrs). But the DM deficient population is replenished by new interactions between dwarfs and massive companions. It is worth noting that our results are robust with respect to the details of the sub-grid recipes implemented in \texttt{NewHorizon}, as this formation mechanism is principally driven by gravitational forces. 

The formation of DM deficient galaxies through tidal stripping of DM, as hypothesised here, is strongly supported by recent observational studies \citep[e.g.][]{Montes2020}, which are located in similar environments to the DM deficient galaxies in \texttt{NewHorizon}. Observations of an excess of GCs in these galaxies, although not directly testable in \texttt{NewHorizon}, also appear to support this formation mechanism (although, as noted above, contributions from other pathways may be required to explain the DM deficient population as a whole).

The results of our study offer a route to identifying dwarf galaxies that may have low DM content, purely from data that is typically available in imaging surveys. Dwarfs that are found close to massive companions with M$_{\star}$ > 10$^{10}$ M$_{\odot}$, at distances less than $\sim$150 kpc, and show stellar tidal features, are likely to be good candidates for being DM deficient systems. The surface brightness of the tidal features produced by such interactions (typically fainter than 29 mag arcsec$^{-2}$) suggests that new and future deep wide surveys, like the Hyper Suprime-Cam SSP or LSST, could be used to identify large samples of such DM deficient dwarf candidates. 

Our study demonstrates that stripping of DM via a tidal interaction, a process that takes place in all environments, can routinely create dwarfs that have DM fractions that deviate significantly from their initial values in the early Universe and, in extreme cases, produces systems that are DM deficient. The existence of such galaxies is therefore an integral feature of galaxy evolution in the standard $\Lambda$CDM paradigm and their existence is not in tension with the predictions of this model. 


\section*{Acknowledgements}
We are grateful to the anonymous referee for many constructive comments that helped us to improve the quality of the original manuscript. RAJ and SK acknowledge support from the STFC [ST/R504786/1, ST/S00615X/1]. SK acknowledges a Senior Research Fellowship from Worcester College Oxford. The research of AS and JD is supported by the Beecroft Trust and STFC. TK was supported in part by the National Research Foundation of Korea (NRF-2017R1A5A1070354 and NRF-2020R1C1C100707911) and in part by the Yonsei University Future-leading Research Initiative (RMS2-2019-22-0216). Some of the numerical work made use of the DiRAC Data Intensive service at Leicester, operated by the University of Leicester IT Services, which forms part of the STFC DiRAC HPC Facility (www.dirac.ac.uk). The equipment was funded by BEIS capital funding via STFC capital grants ST/K000373/1 and ST/R002363/1 and STFC DiRAC Operations grant ST/R001014/1. DiRAC is part of the National e-Infrastructure. This research has used the DiRAC facility, jointly funded by the STFC and the Large Facilities Capital Fund of BIS, and has been partially supported by grant Segal ANR- 19-CE31-0017 of the French ANR. This work was granted access to the HPC resources of CINES under the allocations 2013047012, 2014047012 and 2015047012 made by GENCI and was granted access to the high-performance computing resources of CINES under the allocations c2016047637 and A0020407637 from GENCI, and KISTI (KSC-2017-G2-0003). Large data transfer was supported by KREONET, which is managed and operated by KISTI. This work has made use of the Horizon cluster on which the simulation was post-processed, hosted by the Institut d'Astrophysique de Paris. We thank Stephane Rouberol for running it smoothly for us. 


\section*{Data Availability}

The data that underpins the analysis in this paper were provided by the Horizon-AGN and New Horizon collaborations. The data will be shared on request to the corresponding author, with the permission of the Horizon-AGN and New Horizon collaborations. Local density estimation code is available at \href{https://github.com/garrethmartin/MBEtree}{github.com/garrethmartin/MBEtree}.


\bibliographystyle{mnras}
\bibliography{bib}


\appendix

\section{Orbital distances of dwarfs around their massive centrals}

\begin{figure}
\centering
\includegraphics[width=\columnwidth]{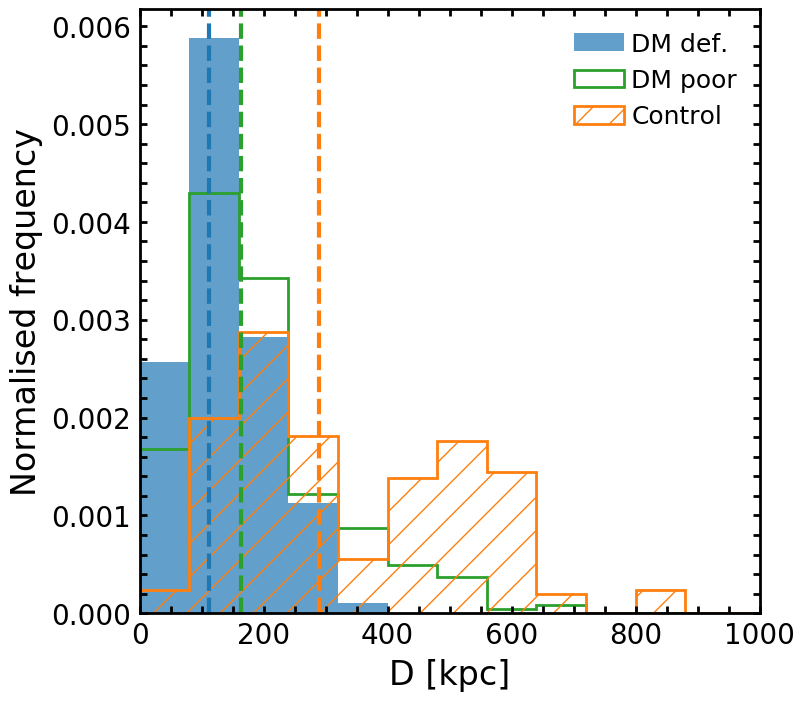}
\caption{Orbital distances of our different satellite populations, in the period after individual dwarfs become satellites through to the end of the simulation. The dashed lines indicate the median value of each distribution.}
\label{fig:dist_nonorm}
\end{figure}

Figure \ref{fig:dist_nonorm} shows the orbital distances of dwarfs around their massive centrals in our different satellite populations. This figure is a version of the right-hand panel of Figure \ref{fig:alldist}, without the distance being normalised by the virial radius of the halo of the massive central. These distances are directly comparable to measured distances between massive galaxies and nearby dwarfs in deep-wide surveys in order to identify DM deficient galaxy candidates. 

\bsp
\label{lastpage}
\end{document}